\documentclass{phc-proc4-auth}

\usepackage{graphicx}
\usepackage{bm}
\usepackage{amssymb}

\begin{document}
\begin{frontmatter}

\title{Doping dependence of superconducting gap in YBa$_2$Cu$_3$O$_{y}$ from 
          universal heat transport}

\author[TOR]{Mike~Sutherland},
\author[TOR]{D.G.~Hawthorn},
\author[TOR]{R.W.~Hill},
\author[TOR]{F.~Ronning},
\author[TOR]{H.~Zhang},
\author[TOR]{E.~Boaknin},
\author[TOR]{M.A.~Tanatar\thanksref{MAT}},
\author[TOR]{J.~Paglione},
\author[UBC]{R.~Liang},
\author[UBC]{D.A.~Bonn},
\author[UBC]{W.N.~Hardy},
\author[TOR]{Louis~Taillefer\corauthref{cor1}\thanksref{LT}}

\address[TOR]{Department of Physics, University of Toronto, Toronto, Ontario, 
Canada M5S 1A7}

\address[UBC]{Department of Physics, University of British Columbia,
Vancouver, British Columbia, Canada V6T 1Z1}

\corauth[cor1]{Corresponding author: Louis.Taillefer@physique.usherb.ca}
\thanks[MAT]{Permanent address: Inst. Surf. Chem., N.A.S. Ukraine}
\thanks[LT]{Current address: Department of Physics, University of 
Sherbrooke, Sherbrooke, Quebec, Canada J1K 2R1.}

\date{\today}

\begin{abstract}
Thermal transport 
in the $T \rightarrow 0$ limit was measured as a function 
of doping in high-quality single crystals of the cuprate superconductor  
YBa$_2$Cu$_3$O$_y$.  
The residual linear term $\kappa_{0}/T $ is found to decrease as one moves 
from the overdoped regime towards the Mott insulator region of the phase diagram.   The 
doping dependence of the low-energy quasiparticle gap extracted from $\kappa_{0}/T $ is 
seen to scale closely with that of the pseudogap, arguing against a non-superconducting 
origin for the pseudogap.  The presence of a linear term for all dopings is evidence against 
the existence of a quantum phase transition to an order parameter with a complex ( $ix$ ) 
component.

\end{abstract}


\end{frontmatter}

In $d$-wave superconductors the presence of nodes in the gap structure leads to the 
existence 
of quasiparticle excitations down to zero energy.  The ability of these excitations to 
transport heat has been shown to be universal ({\it i.e.} independent of impurity scattering )
\cite{Taillefer},
and the residual linear term $\kappa_0/T$ in the thermal conductivity 
$\kappa(T)$ is a direct probe of the low-energy quasiparticle 
spectrum \cite{Durst}.  
At optimal doping in Bi-2212 the agreement between the value 
of the slope of the gap at the node obtained from
$\kappa_{0}/T$ and that measured by ARPES is remarkable \cite{Chiao}.
  
In this paper we investigate the doping dependence of $\kappa_{0}/T$
in the cuprate superconductor YBa$_2$Cu$_3$O$_y$ (YBCO) and find that it decreases monotonically
as one moves from overdoped to underdoped samples.   
The samples used in this study are all high-quality de-twinned single crystals, described 
elsewhere \cite{Sutherland}.   Their $T_c$'s are 62, 44, 93.5 and 
89 K for oxygen dopings $y = 6.54$, 6.6, 6.95 and 6.99, respectively.
Thermal conductivity measurements were made in a dilution 
refrigerator down to 40 mK, with current along the a-axis.

\begin{figure}[tbh]
\centering
\resizebox{\columnwidth}{!}{\includegraphics[angle=0]{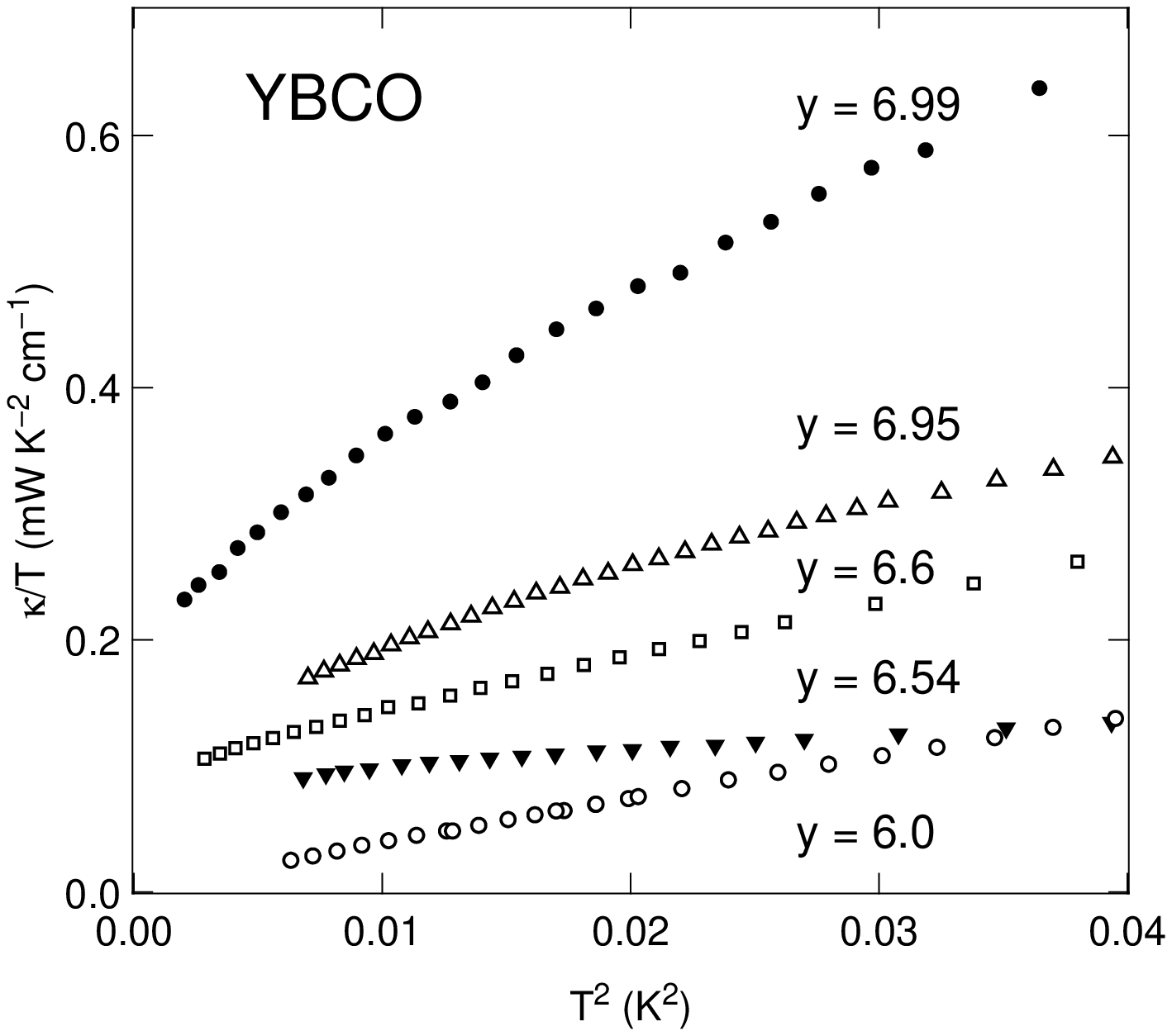}}
\caption{\label{fig:kappa}
Thermal conductivity of $\rm YBa_2Cu_3O_{y}$.}
\end{figure}

In Fig.~\ref{fig:kappa}, the thermal conductivity of each sample is plotted 
as $\kappa/T$ vs. $T^{2}$.  To seperate out the electronic from the phononic contributions, 
we model $\kappa$ to be of the form $\kappa = \kappa_\mathrm{el} + \kappa_\mathrm{ph} 
= A T +  B T^{\alpha }$.  The $T$-linear electronic term is a well known result of 
$d$-wave superconductors \cite{Durst}, while the $T^{\alpha}$ term is an empirical fit to 
the phonon conductivity at low temperatures \cite{Sutherland}. 

It is clear from Fig.~\ref{fig:kappa} that the value of the linear 
term decreases monotonically as one moves from the slightly overdoped $y=6.99$ sample to 
the strongly underdoped $y=6.54$ sample (It becomes zero in the non-superconducting
parent compound $y=0$.) 
The presence of such a 
linear term for all dopings excludes the possibility of a quantum 
phase transition from a pure $d$-wave form to a $d+ix$ state suggested by some authors,  
since the effect of such a transition would be to gap out the quasiparticle 
spectrum and hence eliminate any residual linear term. 

From our measurements of $\kappa_0/T$ we can extract a measure of the 
quasiparticle gap in the following manner.   
The quasiparticle thermal conductivity of a $d$-wave superconductor at $T \to 0$ is given by \cite{Durst}:

\begin{equation}
 \frac{\kappa_0}{T} = \frac{k_B^2}{3 \hbar} \frac{n}{d} \left( \frac{v_F}{v_2} + \frac{v_2}{v_F} \right),
 \label{eq:koT}
\end{equation}

where $n$ is the number of $\rm{CuO_2}$ planes per unit cell and $d$ is the $c$-axis 
lattice constant.  Here $v_F$ and $v_2$ are the quasiparticle velocities normal and 
tangential to the Fermi surface at the node, respectively.  
From ARPES measurements in cuprates, we note the the value of 
$v_{F}$ is doping independent over a broad range of dopings \cite{Mesot,Shen} and thus a 
study of the doping 
dependence of $\kappa_{0}/T$ is a study of the doping dependence of $v_{2}$, 
where $v_{2}$ is simply $\frac{1}{\hbar k_F}\frac{d\Delta}{d{\phi}}\Big|_{node}$, the 
slope of the superconducting gap at the nodes.  In order to compare with measures of the 
gap maximum $\Delta_0$, we assume 
the pure $d$-wave form $\Delta = \Delta_0 {\rm cos}2\phi$ throughout the doping phase 
diagram, so that 
$\Delta_0 = \hbar k_F v_2 / 2$.  In Fig. ~2 we plot the value of $\Delta_0$ vs. hole 
concentration $p$, estimated from $T_c$ \cite{Sutherland}.
The value of the energy gap maximum in Bi-2212, determined by ARPES 
measurements in the superconducting state \cite{campuzano} 
and the normal state \cite{norman,white,loeser}, is shown alongside our own data.  
For comparison, a plot of 
the weak-coupling BCS expectation, $\Delta_{\rm BCS}=2.14 k_B T_c$, is also displayed.

\begin{figure}[htb]
\centering
\resizebox{\columnwidth}{!}{\includegraphics{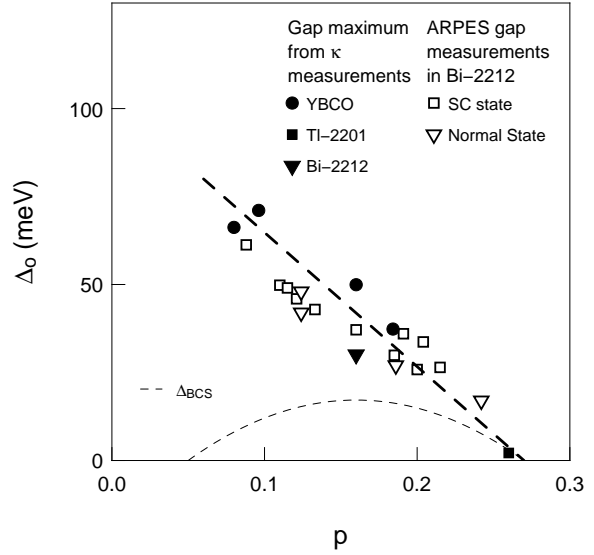}}
\caption{\label{fig:kvsH}  $\Delta_0$ obtained from our measurements of $\kappa_0/T$ (see text) on YBCO,
as well as previous measurements on Bi-2212 (Ref.~3) and Tl-2201 (Ref.~11) 
(filled symbols). 
For comparison, a BCS gap of the form $\Delta_{\rm BCS}=2.14 k_B T_c$ is also plotted.  
The value of the energy gap maximum in Bi-2212, as determined by ARPES, 
is shown as measured in the superconducting state \cite{campuzano} and the normal state 
\cite{norman,white,loeser}  (open symbols).  The thick dashed line is a guide to the eye.}
\end{figure}

Despite a remarkable 
quantitative agreement with BCS theory in strongly overdoped cuprates \cite{Proust},
the value of the gap measured in the bulk at $T = 0$ does not follow the trend 
expected from BCS theory in the underdoped regime, where $\Delta_0$ should scale with $T_c$.  
Rather, the gap seen by low energy quasiparticles follows closely the energy scale set by the 
pseudogap.   This similarity in scaling points to a common origin, which allow us to say 
the following things on the nature of the pseudogap.  First, due to the very existence of 
a residual linear term, the (total) gap seen in thermal conductivity at $T \to 0$ is one 
that must have nodes. Secondly, it has a linear dispersion as in a $d$-wave gap ({\it i.e.} 
it has a Dirac-like spectrum).  Thirdly, it is a quasiparticle gap and not just a spin gap.  
In summary these observations argue strongly for a superconducting origin to the pseudogap, 
and our measurements
also rule out the possibility of a bulk order parameter of the type $d+ix$,
appearing at a putative quantum phase transition.

\vspace{8pt}

This work was supported by the Canadian Institute for Advanced Research 
and NSERC of Canada.


\bibliographystyle{elsart-num}
\bibliography{M2Sbib}

\end{document}